\documentclass[onecolumn,secnumarabic,amssymb, nobibnotes, aps, preprint, prd]{revtex4-1}

\usepackage{graphicx}
\usepackage{bm}

\begin{document}

\title{Quantum-mechanical treatment of $^{1}$H$^{+}$ / $^{2}$H$^{+}$ / $^{3}$H$^{+}$ ion dynamics in carbon nanotubes}%
	
\author{Anthony B. Kunkel, Daniel J. Finazzo, Renat A. Sultanov, and Dennis Guster, }%

\begin{abstract} 
An investigation of quantum dynamical properties was conducted for $^{1}$H$^{+}$ / $^{2}$H$^{+}$ / $^{3}$H$^{+}$ ions as they traverse through a Carbon nanotube (CNT). The investigation is focused on a Fortran based program simulating a hydrogen ion passing through a CNT. Our work included testing convergence of the simulations, understanding dynamical behavior of hydrogen isotopes, and investigating the applied H$^{+}$ + CNT potential energy curve. 
\end{abstract}
\maketitle

\section{Introduction}
Hydrogen is known to be a key element that can be used in alternative energy. The alternative energy field is a extremely fast growing research topic since our current main energy source, fossil 
fuels, will not last forever. Hydrogen has been seen to benefit the alternative energy field by being 
an abundant and an efficient source of energy. Although, there are some hurdles that exist with 
using hydrogen as a fuel source, specifically in the act of storing it \cite{Sharma20151151,Im20122749,Orhan2015801}.

Hydrogen is rarely found alone and is most often found bonded to other elements. This is an solvable 
problem since it is relatively easy to dissociate hydrogen from water molecules and separate them from the oxygen \cite{Pletcher201115089}. Another problem subsequently arises, considering that hydrogen is extremely 
dangerous to store due to its combustible properties \cite{Pasman20112407,Petukhov20095924}. This problem is one of the main reason using 
hydrogen as a fuel source has been at a standstill. It is now of high importance to find a medium that 
will store and release hydrogen at will. One possible medium that is under investigation in this 
research and many others is using carbon nanotubes (CNT)\cite{Hynek1997601,PhysRevB.67.245413,4287144620090601,Skouteris201318,0953-8984-14-17-201}. If stored properly hydrogen could not only be stored 
safely but in a very small and organized space \cite{RafiiTabar2004235}. 

Another interesting use of the CNT is the potential for using it as an isotopic filter \cite{PhysRevA.64.022903,PhysRevB.63.245419,PhysRevLett.82.956,PhysRevB.65.014503}. The work found when hydrogen isotopes traveled through a single-walled carbon nanotube (SWNT) displayed interesting results for isotopic separation. Since hydrogen isotopes vary in kinetic energy, due to their mass difference, they will exit the SWNT at different times \cite{doi:10.1021/nn901592x,doi:10.1021/jp054511p,Ghosh201547,PhysRevLett.94.175501,doi:10.1021/jp030601n,doi:10.1021/ja0502573}.

The interesting aspects investigated in this paper, is how the separation changes when varying the parameters of the CNT such as size and/or temperature. The larger the isotopic seperation shown would be incredibly useful in fields where hydrogen isotopes are of great importance. For instance, deuterium is sometimes found bonded to oxygen in the form of heavy water which must be filtered separately from hydrogen. These deuterium atoms can then be used in research with tritium for tests in the nuclear field \cite{edsgcl.36234808920140101}. Therefore, it is critical to find inexpensive ways to separate all of these isotopes of hydrogen. CNT might be the best solution for this low-cost isotopic separation \cite{doi:10.1021/jp064745o}.

 In this experiment a Fortran based program called 
CYLWAVE is used to find the flux of outgoing hydrogen atoms from the CNT. We are focused on 
manipulating the CNT to represent a physical system that takes hydrogen in but does not allow it to 
escape. Another focus is on how the isotopic separation is affected by the variables of the CNT. We are also interested in visually simulating the potential energy curve to get an understanding on what potentials are applied to the particles inside our CNT
\cite{Skouteris2009459}.

In this paper Sec. II describes the mechanics of the CYLWAVE program. It also explains how our variations of parameters in the program were approached. Sec. III gives an overview of information obtained through the course of our research. Sec. IV reviews the information presented throughout this document. Future work and outlook is also given in this section. All units used are in atomic units unless otherwise stated. 

\section{Methods}
The program as mentioned before, utilizes a wavepacket propagation approach inside a stationary CNT. Since the problem at hand is best suited in cylindrical coordinates it is necessary to deal with the Hamiltonian operator in cylindrical coordinates as well. The Hamiltonian then becomes as follows with $r$ as the radius, $\theta$ as the azimuthal angle, and $z$ as the longitudinal distance:  
\begin{equation}
\hat{H}=-\frac{\hbar^{2}}{2m}\left( \frac{\partial^{2}}{\partial z^{2}}+\frac{\partial^{2}}{\partial r^{2}}+\frac{1}{r}\frac{\partial}{\partial r}+\frac{1}{r^{2}}\frac{\partial^{2}}{\partial \theta^{2}}\right) +V(r,\theta,z).
\end{equation}
This equation may be further simplified by dealing with the first order derivative in r;
\begin{equation}
\hat{H}=-\frac{\hbar^{2}}{2m}\left( \frac{\partial^{2}}{\partial z^{2}}+r^{-1/2}\frac{\partial^{2}}{\partial r^{2}}r^{1/2}+\frac{1}{4r^{2}}+\frac{1}{r^{2}}\frac{\partial^{2}}{\partial \theta^{2}}\right) +V(r,\theta,z).
\end{equation}
Finally, if the time-dependent wavefunction $\Psi$ is replaced with $r^{1/2}\psi$ the Hamiltonian operator becomes \cite{1432}:

\begin{equation}
\hat{H}=-\frac{\hbar^{2}}{2m}\left( \frac{\partial^{2}}{\partial z^{2}}+\frac{\partial^{2}}{\partial r^{2}}+\frac{1}{4r^{2}}+\frac{1}{r^{2}}\frac{\partial^{2}}{\partial \theta^{2}}\right) +V(r,\theta,z).
\end{equation}

The function $V$ is a combination three different potential functions, to accurately design the potential energy surface. The sum of Lennard-Jones (6-12) potentials have been included to account for the interaction between the hydrogen atom and the carbon atoms. The Lennard-Jones potential is described as: 
\begin{equation}
V=4\varepsilon\left[\left(\frac{\sigma}{r}\right)^{12}-\left(\frac{\sigma}{r}\right)^{6}\right].
\end{equation}
Here $\varepsilon$ is the potential well depth and $\sigma$ is the distance where the inter-particle potential is zero.

There also are two simulated potentials designed to exclude nonphysical characteristics of the wavepacket. One of these potentials is placed at the longitudinal end of the tube in order to absorb the exiting particle. The other is located at the radial edge of the CNT to allow for an exponential decay of the wavefunction outside of the nanotube, to account for possible tunneling effects. 

 
The program allowed us to change the parameters of the particle and the CNT using the input, 
parameter, and potential energy surface files. Our experiment was solely focused on working with 
the input file in the interest of only changing specific variables and achieving convergence with our 
results. With the input file we were able to change the particle’s mass and longitudinal energy. It 
also allowed for changes to the components of the CNT such as its $z$, $r$, and $\theta$ 
integration points. 
 
After understanding the mechanics of the program it was a logical next step to find which 
components of the input file would be varied and which need to be held constant. The original file was designed as a 
test simulation to ensure that the program was functioning properly. The test simulation was found to 
be working correctly but it lacked detailed results, especially in the number of azimuthal integration 
points. We solved this problem by holding all parameters constant and changing the number of 
azimuthal points until convergence was found. This same technique was used in finding convergence in the integration points for the radial and longitudinal coordinates. 
  
The final steps in our simulation were to manipulate the longitudinal energy to determine the 
possibility of separation. The wavepacket approach breaks down at low energy and limited our ability to 
do simulations at low temperatures, i.e., less than 50 Kelvin (K). Although we are curious on the possibility of separation at temperatures near room temperature (300 K).

We also took time to investigate other isotopes of hydrogen. Deuterium and tritium were used since they might also be traveling through a particular CNT. The original mass of the particle was 1836 atomic units (a.u.), which corresponded to a hydrogen ion. The mass was doubled and 
tripled, respectively, to simulate deuterium and tritium.

\section{Results}
The Lennard-Jones potential is of great importance to the transportation of the hydrogen ion, and is imperative that we take time investigating how the program simulates the potential. We were able to visualize the potential by programming the simulation to display the values of the potential and the r coordinates, while keeping the z and theta values static. This approach utilizes the symmetry of the potential, understanding that the potential is dependent on r but remains constant for varying values of both $\theta$ and $z$. These potential values were then plotted in Figure 1 to display the potential energy curve used in the program. As mentioned before, there is a radial potential designed to restrict the particle to only be able to tunnel out of the CNT. This potential has been omitted since it is not part of the Lennard-Jones potential.
  
It was apparent that low values of integration points was not sufficient in getting detailed results from the program. The 
values were tested for points 8 through 16. There were no visible changes in flux for azimuthal 
points 10 through 16, as shown in Figure 1, so it was concluded that convergence had been found. 
For the future runs we held the azimuthal points at 10 to save time on computation.
 
This particular approach was applied to find convergence in the r and z coordinates. The initial number of integration points were 144 and 72 for z and r, respectively. The z-points were increased to 216 and 432 while keeping all other parameters constant. The results in Figure 3  show that there is no change in flux due to the increase in integration points. Similarly we increased the integration points of the r coordinate to 144. The results from these simulations, Figure 4, portrayed that there was no change between the changes r-points. Given this data we can see that the program is convergent for each of our chosen values of integration points. Although, there was a notable increase in computation time as we changed the integration points. Therefore we choose to use 72 r-points and 216 z-points to shorten the time per simulation.
   
Another interesting result is due to the findings of the flux data associated with the particles 
hydrogen, deuterium, and tritium. Simulations were run on the three particles using the same values 
in all parameters except for the particles mass. Figure 5 shows that the particles vary in the amount 
of flux and at what time the maximum flux exits the tube. It is important to note that the flux is shown to be negative in the results due 
to the convention of the flux operator in the program, which is given by the equation 
\begin{equation}
\hat{F}=\frac{i}{\hbar}\left[\hat{H},\theta(z-z_{0})\right].
\end{equation}
With $\theta(z-z_{0})$ being the Heaviside step function. This means that the results actually correspond to a maximum 
as opposed to the appearance of a minimum. The figure shows that hydrogen and 
its isotopes in fact may exit the CNT at different times. The problem that arises is the flux of each
particle are overlapping.

The temperature was changed to determine its affect on the amount of overlap between the particles. The temperature was varied from 50 K to 300 K as shown in Figures 5-9. The overlap persist with the change in temperature. Although, it does show interesting results in the change in flux. As the temperature is decreased we can see that the maximum flux continues to decrease for each particle. It can also be seen that the system with the least energy also has maximum flux at a later time than the higher energy graphs. These trends may be described by the particles change in velocity as the temperature is varied. We might also assume that the flux is also proportional to the velocity.   

\section{Conclusion}
In the experiment we were able to test the CYLWAVE program for convergence of the hydrogen ion inside the CNT. We tested several areas of the program including; the integration points for $\theta$, $r$, and $z$. We also investigated the change in flux in correspondence to the changes in longitudinal energy. Convergence was found when using integration points at 10, 216, and 72 for $\theta$, $r$, $z$ respectively. It was also found that the CNT produced a separation between peak flux for isotopes of hydrogen. The separation was seen to remain nearly constant as the energy modified from high to low. This fact requires more inspection in future work. 

As this work continues it will be important to look at several more aspects of the program and to adapt the CNT itself. One area of interest will be focused on the Lennard-Jones potential. More time must be required to truly understand the potentials affect of the CNT and on the transported particle. The potential could be modified to better suit other particles other than hydrogen ions and isotopes. Another concern that must be continued from this paper is the investigation of isotope separation. A closer look would shed light on how the size of CNT affects this separation. More specifically, how the change in radius and length adjusts the separation between the peaks in outgoing flux. These changes may also give way to finding a way to greatly decrease the outgoing particle's flux. If large decreases in the flux can be shown it would greatly benefit the hydrogen storage and energy fields.

 \bibliography{science1.bib}
 	
 \bibliographystyle{ieeetr}
\newpage
 
\begin{figure}	
	\centering
	\includegraphics[width=160mm, scale=0.90]{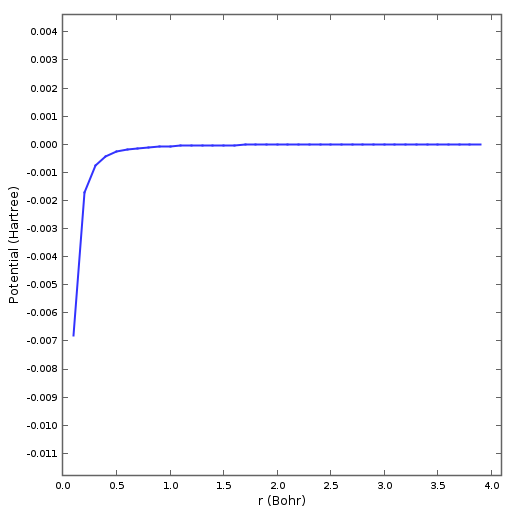}
	\caption{Plot of the Lennard Jones potential applied to the H$^{+}$+CNT.}
\end{figure}

\begin{figure}	
\centering
\includegraphics[scale=0.75]{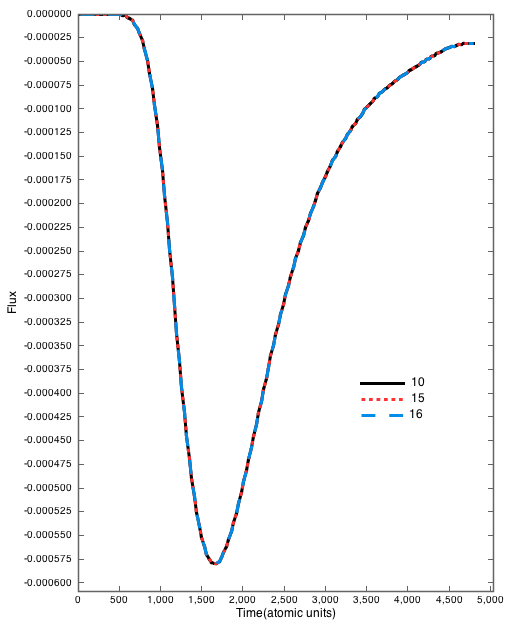}
\caption{Convergence of azimuthal points 10, 15, and 16.}
\end{figure}

\begin{figure}	
	\centering
	\includegraphics[scale=0.75]{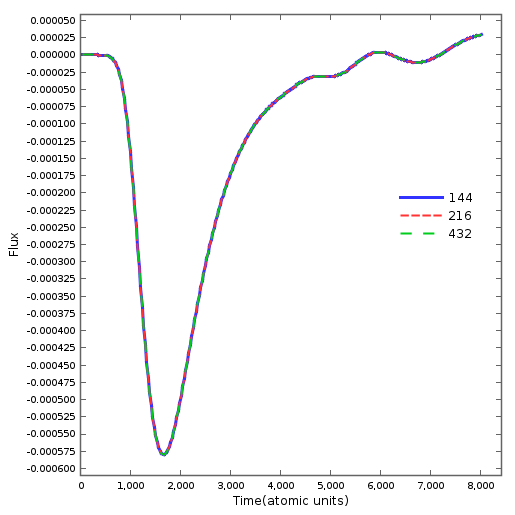}
	\caption{Convergence of z-points 144, 216, and 432.}
\end{figure}

\begin{figure}	
	\centering
	\includegraphics[scale=0.75]{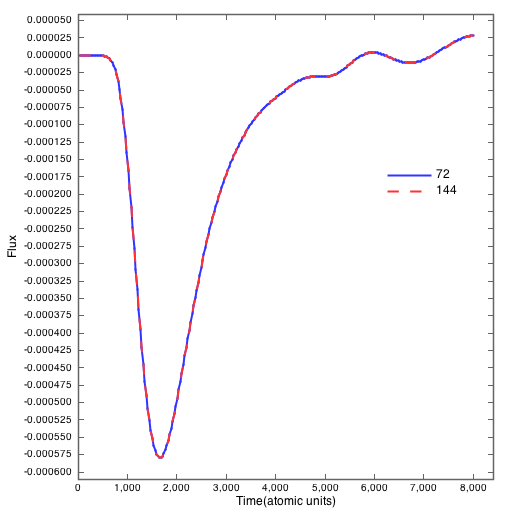}
	\caption{Convergence of r-points 72 and 144.}
\end{figure}

\begin{figure}	
	\centering
	\includegraphics[scale=0.65]{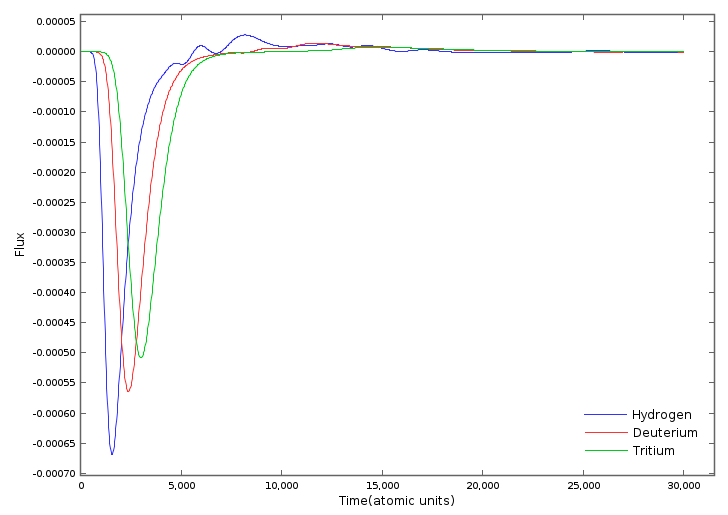}
	\caption{Flux output of hydrogen, deuterium, and tritium at 300 K.}
\end{figure}

\begin{figure}	
	\centering
\includegraphics[scale=0.75]{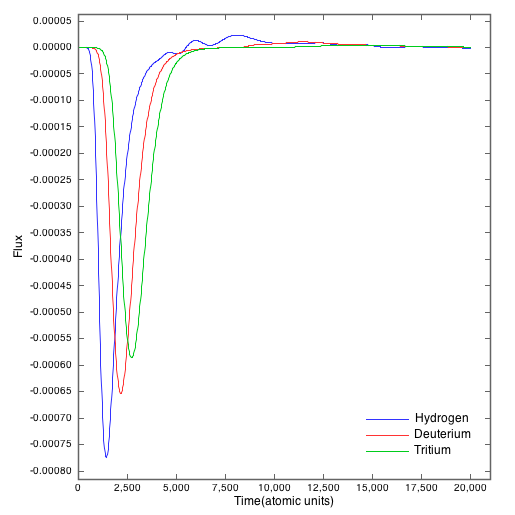}
\caption{Flux output of hydrogen, deuterium, and tritium at 200 K.}
\end{figure}

\begin{figure}	
	\centering
	\includegraphics[scale=0.75]{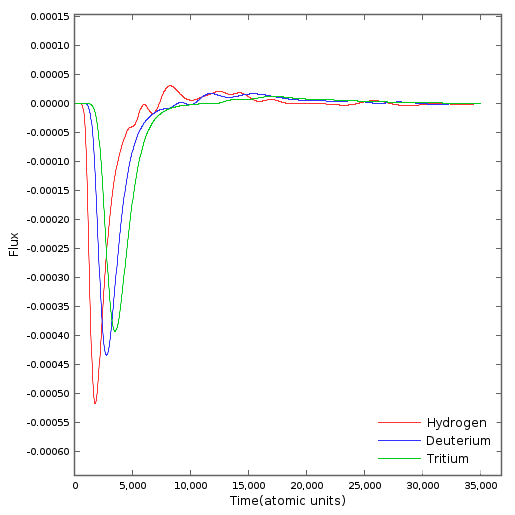}
	\caption{Flux output of hydrogen, deuterium, and tritium at 150 K.}
\end{figure}

\begin{figure}	
	\centering
	\includegraphics[scale=0.75]{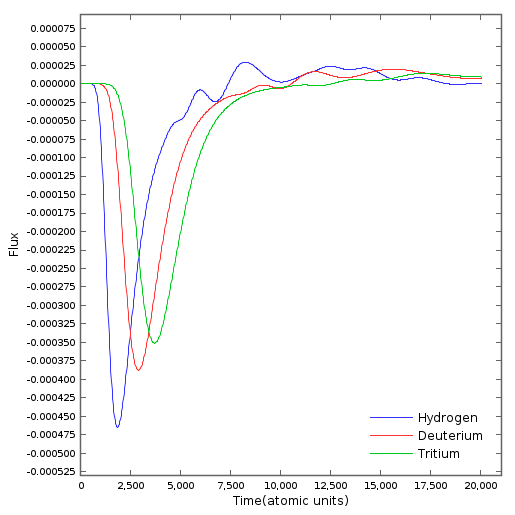}
	\caption{Flux output of hydrogen, deuterium, and tritium at 100 K.}
\end{figure}

\begin{figure}	
	\centering
	\includegraphics[scale=0.75]{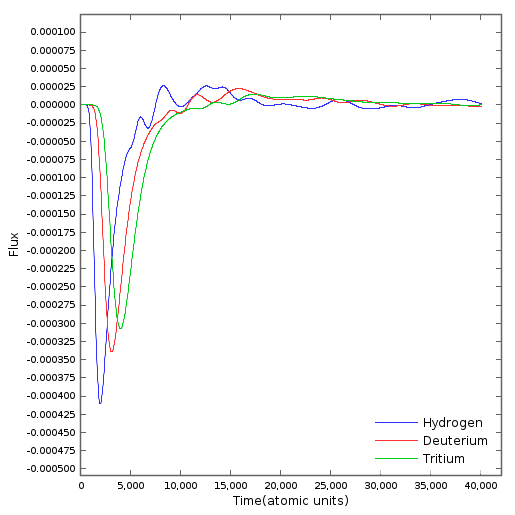}
	\caption{Flux output of hydrogen, deuterium, and tritium at 50 K.}
\end{figure}

\end{document}